# Designer-Driven Topology Optimization for Pipelined Analog to Digital Converters

Yu-Tsun Chien[†], Dong Chen, Jea-Hong Lou[†], Gin-Kou Ma[†], Rob A. Rutenbar, and Tamal Mukherjee

†SoC Technology Center, Industrial Technology Research Institute, Hsinchu, Taiwan
and ECE Department, Carnegie Mellon University, Pittsburgh, Pennsylvania, USA

## Abstract

*This paper suggests a practical "hybrid" synthesis methodology which integrates designer-derived analytical models for system-level description with simulation-based models at the circuit level. We show how to optimize stage-resolution to minimize the power in a pipelined ADC. Exploration (via detailed synthesis) of several ADC configurations is used to show that a 4-3-2... resolution distribution uses the least power for a 13-bit 40 MSPS converter in a 0.25 µm CMOS process.*

## 1. Introduction

In the digital world, RTL/logic synthesis is an indispensable tool for allowing system designers to explore high-level architecture, functional partitioning, and performance tradeoffs. Synthesis serves the role of *completing* a block to a level of concreteness that can be used to properly evaluate the merits of architecture-level tradeoff decisions. The questions we address in this paper are how recently introduced commercial-quality analog synthesis tools (e.g., [1][2]) might play the same role in system-level analog design, and what a systematic methodology for this sort of design might look like. Two broad approaches have been suggested:

- *Equation-based* methods avoid simulation entirely, and strive to represent the design at all levels in an analytical form [3-5]. Convex models are the most successful here [4-5]. However, the method makes some serious tradeoffs emphasizing speed at the cost of design accuracy.
- *Hierarchical simulation-based methods* use macromodels for the system-level design, and detailed models for the basic blocks, and then link these numerically. [6] is perhaps the most aggressive attempt in this direction. An alternative approach builds *Pareto* tradeoff curves for each basic block via detailed synthesis [7-9], then uses these to "parameterize" a system-level model. Both techniques are attractive, but not yet well-supported in commercial synthesis tools.

We suggest a practical "hybrid" approach which integrates well with the current crop of commercial synthesis tools, and is moreover consistent with the style in which most analog system designers prefer to work. To make this concrete, we look at system-level architecture/topology *power minimization* for 40 MSPS pipelined ADCs in a 0.25µm 3.3V CMOS process with resolutions from 10 to 13 bits.

## 2. Candidate Enumeration

The pipelined ADC consists of a front-end sample and hold amplifier (S/H amplifier) and $M$ pipelined stages and the number of bits to be converted in each stage ($m_i$). One extra bit from each stage is used to implement digital correction logic. Designers have used a variety of configurations $m_1$-$m_2$-$m_3\cdots$ to design pipelined ADCs, starting from the classical 2-2-2$\cdots$ or 1.5 bits per stage topology [5] to the recent 4-2-2$\cdots$ [10]. The possible configuration sets $\{m_1$-$m_2$-$m_3,\cdots\}$ for a K-bit pipelined ADC is governed by

$$\sum_{i=1}^{M}(m_i - 1) = K$$

where K is the total effective number of bits over M stages with digital correction circuitry.

In this paper, candidate enumeration is used to explore all possible configurations such that $m_i \leqq 4$ and $m_i \geqq m_{i+1}$. The $m_i \leqq 4$ constraint is due to closed-loop bandwidth concerns. The $m_i \geqq m_{i+1}$ constraint arises because of the area factor and is often used implicitly [5][10]. Additionally, we only consider the first few stages such that the output resolution exceeds 7 bits. This is because ADC power is mainly consumed by the starting few bits [5]. These reduce the design space complexity to a manageable enumerated set of seven different candidates. Each candidate has *several* MDAC stages to be synthesized using method in section 3. The MDAC block-level specifications can be translated from the ADC system-level specifications and the value $m_i$ for the enumerated candidate

## 3. Block-level Synthesis Design Flow

The proposed block-level synthesis design flow combines circuit analysis with simulation to reduce the design space and speed up transistor-level evaluation, enabling use of commercial cell-level synthesis tools. First, once the circuit topology of the MDAC block is decided, Driving-Point Impedances (DPI) / Signal-Flow Graphs (SFG) is used to draw the signal flow graph equivalent of the circuit. Second, the circuit symbolic transfer function is derived from the SFG by using Mason's rule. With SFG and symbolic transfer function information, circuit characteristics such as poles/zeros, gain, phase-margin, are analyzed. The range of the design variables that define the design space and the design constraints are reduced using the DPI/SFG analysis results.

When circuits experience large dynamic swing, simulation-based evaluation produces trustworthy results within a short period of time. When circuit behavior is linear, transfer functions based on small signal parameters evaluate circuit performances accurately and efficiently. Combining these approaches has the advantage of high simulation accuracy and fast equation evaluation. Thus, evaluation of each candidate solution involves: 1) DC simulation to extract small signal values, 2) formulating the numerical transfer function, and using the toolkit for hybrid equation+simulation evaluation. This evaluation procedure is performed automatically in each synthesis iteration.

## 4. Topology Optimization Results

Eleven MDACs used to enumerate the seven 13-bit ADC configurations were synthesized (using the Cadence NeoCircuit tool). Fig. 1 shows that the power of the first stage is mostly independent of the resolution of the first stage. Choosing 4-bits in the first stage, which minimizes the bits in the other stages, optimizes the power in the 13-bit case.

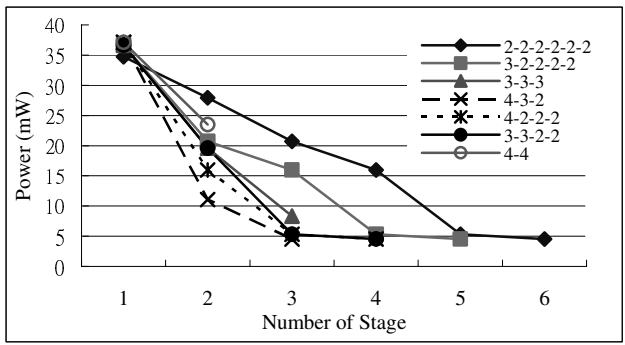

**Fig. 1 Stage power for 13-bit ADC configuration**

MDAC power is added to the sub-ADC comparator power to obtain the overall ADC power. Fig. 2 presents the total power for the stages with output resolution exceeding 7 bits in each of the enumerated architectures for the 10~13-bit pipelined ADCs. 3-2…, 4-2…, 4-2-2…, 4-3-2… are the optimum candidate numeration for 10, 11, 12 and 13-bit, respectively. 2-bit at the last stage is the common optimum candidate numeration for 10~13-bit. More data raises some instructive implications that can be summarized in Fig. 3.

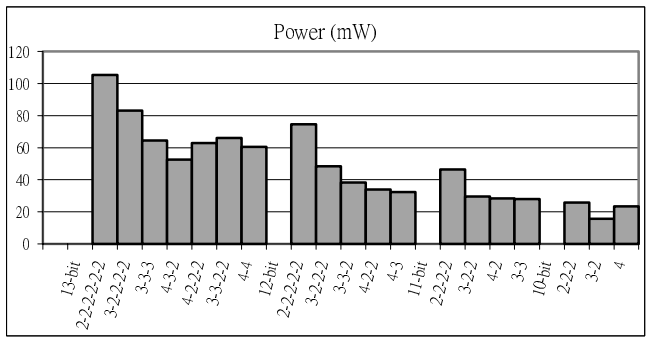

**Fig. 2 Total power for first 6 bits of pipelined ADC**

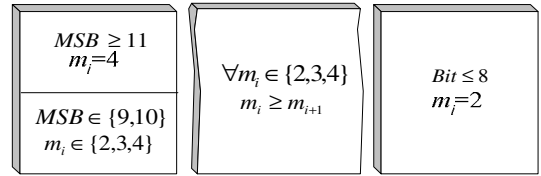

**Fig. 3 Pipelined ADC optimum candidate enumeration**

Setting up the first synthesis required 2-3 weeks, however, the time reduced dramatically to 1 day for subsequent blocks, which only involve retargeting of specifications. In contrast, manual design for each block costs a designer 1-2 weeks in our experience, depending on performance requirements. This use of enumeration and cell synthesis with fast evaluation is suitable for system level optimization of circuits that are composed of similar blocks with varying specifications.

## 5. Conclusion

Existing analog synthesis methodologies are limited by their ability to scale to larger circuits. A system-level synthesis method which enumerates block architecture alternatives, and then performs block-synthesis using a hybrid of simulation-based and equation-based evaluation overcomes this limitation. Enumeration of MDAC synthesis results for optimizing pipelined ADC power is used to demonstrate the efficacy of this method.

### Acknowledgements
The authors would like to thank the Industrial Technology Research Institute Laboratory @ Carnegie Mellon for financial support and Neolinear/Cadence for access to NeoCircuit®.